# 1150 year long ice core record of the Ross Sea Polynya, Antarctica

Thomas Manuel Beers[1,2], Paul Andrew Mayewski[1,2], Andrei Kurbatov[1,2], Daniel Dixon[1], Nancy Bertler[1,3], Sean Birkel[1], Kirk A. Maasch[1,2], Michael Handley[1], Jeff Auger[1,2], Thomas Blunier[4], Peter Neff[3], Andrea Tuohy[3] and Marius Folden Simonsen[4].

1. Climate Change Institute, University of Maine
2. School of Earth and Climate Sciences, University of Maine
3. Victoria University of Wellington and GNS Science
4. University of Copenhagen

**Knowledge of the past behavior of Antarctic polynyas such as the Ross and Weddell Seas' contributes to the understanding of biological productivity, sea ice production, katabatic and Southern Hemisphere Westerly (SHW) wind strength, Antarctic bottom water (ABW) formation, and marine $CO_2$ sequestration. Previous studies link barium (Ba) marine sedimentation to polynya primary productivity (Bonn et al., 1998; McManus et al., 2002; Pirrung et al., 2008), polynya area to katabatic wind strength and proximal cyclones (Bromwich et al., 1998; Drucker et al., 2011), and highlight the influence of Ross Ice Shelf calving event effects on the Ross Sea Polynya (RSP) (Rhodes et al., 2009). Here we use the RICE ice core, located just 120 km from the Ross Ice Shelf front to capture 1150 years of RSP behavior. We link atmospheric Ba fluctuations to Ba marine sedimentation in the summer Ross Sea Polynya, creating the first deep ice core based RSP proxy. RSP area is currently the smallest ever observed over our 1150-year record, and varied throughout the Little Ice Age with fluctuations in Amundsen Sea Low (ASL) strength related with Ross Sea cyclones. Past RSP reconstructions allow us to predict future responses of RSP area to future climate change, which is of special interest considering the recent disappearance of the Weddell Sea Polynya in response to anthropogenic forcing (de Lavergne et al., 2014).**

The Roosevelt Island Climate Evolution (RICE) ice core (figure 1 and supplementary figure 5) is ideally located (79.36398 S, 161.70645 W, 509 m AMSL) to capture Ross Sea Polynya fluctuations. It was drilled 120 km from the Ross Ice Shelf calving front, ~350 km coastward of the Siple Dome deep ice core and ~1000 km coastward of WAIS Divide ice cores. The glaciochemical record is marine dominated, confirmed by Empirical Orthogonal Function (EOF) calculations (see methods and supplementary table 1).

We preformed correlations looking for the strongest association between our element record and climate variables. Correlations between annual means of all elements to ERA-Interim climate reanalysis variables proved Na, S and Ba correlations to be the strongest associations (see methods for further description). Annual average and maximum Na and S compared to December-January-February (DJF) sea ice concentration (SIC) yields a 0.6 correlation in the Ross Sea sector consistent with

Climate Change Institute | University of Maineignoreprevious studies (Sneed et al., 2011). However, Ba annual average concentrations offer the strongest correlation with ERA Interim DJF SIC (up to r = 0.8) over the area of the RSP (figure 1).  We suggest that the RICE Ba proxy for the RSP is a consequence of polynya expansion via atmospheric conditions (katabatic wind surges, proximal low-pressure systems, SHW wind strength) and sea ice-ocean interactions that influence austral summer biogenic blooms, ABW formation and Ross Sea $CO_2$ sequestration. Further, understanding of past fluctuations in polynya area may provide a key to predict future changes in such environments. In developing the RSP proxy we built upon previously suggested mechanisms. We attempted to correlate DJF Ba concentrations with RSP sea ice, but found weaker correlations most likely because of other unknown sources of Ba during DJF, such as terrestrial dust, that would be overpowered in annual average Ba. Annual average Ba correlations are the strongest most likely because of the intense Ba marine sedimentation occurring during DJF (Bonn et al., 1998; McManus et al., 1998; Pirrung et al., 2008), as Ba marine sedimentation does not occur during other seasons due to limited or no sunlight in the Ross Sea Embayment. In the following text we describe this mechanism in detail.

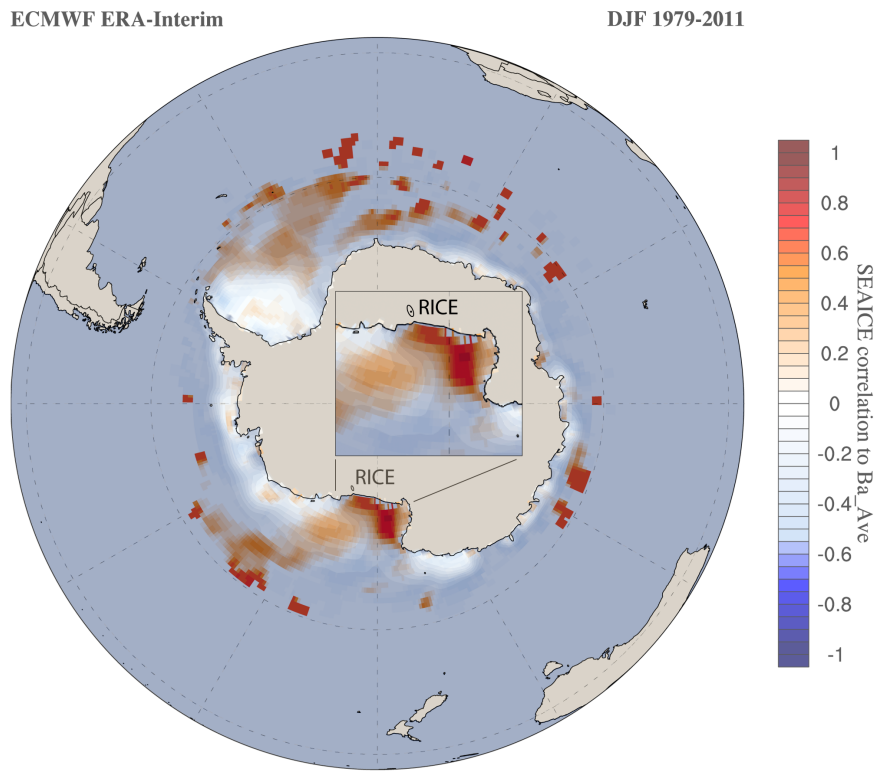

Figure 1: RICE Ba time series correlation to DJF sea ice concentration derived from ERA Interim climate reanalysis data. The Ross Sea Embayment is enlarged (overlying the Antarctic continent) with average DJF sea ice concentration plotted in the contour layer below. Notice that the correlation not only reaches a maximum at 0.8, but also captures the shape of the DJF RSP.  Image obtained using Climate Reanalyzer (http://cci-reanalyzer.org), Climate Change Institute, University of Maine, USA.

As the RSP opens during the austral summer (DJF) and remains open by katabatic wind surges and proximal frequent low-pressure systems north of Roosevelt Island (Bromwich et al., 1998; Drucker et al., 2011), sunlight penetrates the water column fueling a biogenic bloom (figure 2). As previous studies find (Bonn et al., 1998; McManus et al., 2002; Pirrung et al., 2008), Ba sulfate is deposited in marine sediment as organics decay during the biogenic bloom. Katabatic winds keep the RSP open by pushing newly formed sea ice offshore and at the same time entrain Ba sea spray from the RSP. The persistent proximal low-pressure system just north of RICE (supplementary figure 6) recirculates sea spray back onto the continent making sea spray available as condensation nuclei, subsequently depositing it at RICE. The sea spray is depleted in Ba due to Ba sedimentation in DJF relative to other seasons without biogenic influence, i.e. the austral fall, winter and spring. Fluctuations in sea ice extent in the RSP area are integral to RSP primary productivity through sunlight blocking, thus more (less) sea ice, less (more) Ba sulfate marine sedimentation, more (less) Ba atmospheric concentration available for transport to RICE during DJF.

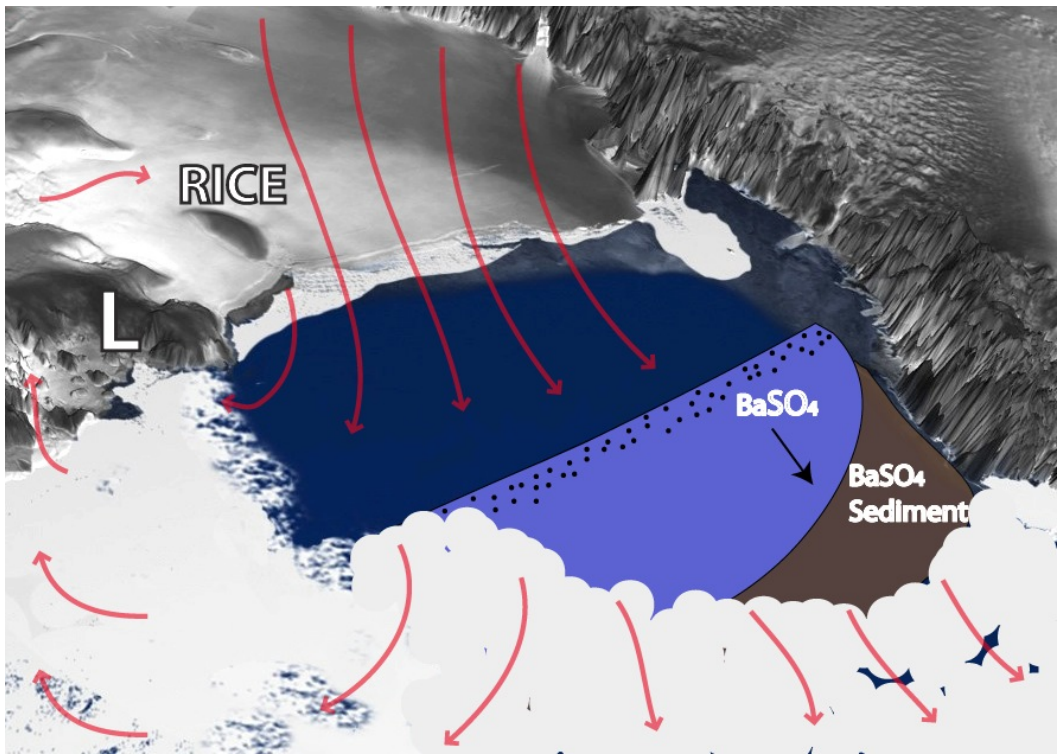

Figure 2: Idealized mechanism for Ba depletion during the DJF polynya. Image modified from NASA IceBridge (http://www.nasa.gov/mission_pages/icebridge). Idealized katabatic winds are shown as red arrows (wind direction and strength). "L" denotes an idealized low-pressure system.

The RICE Ba time series (figure 3) also reveals the influence of Ross Sea iceberg calving events B-9, B-15, B-17 and C-18 and C-19. Sharp increases are recorded in Ba concentrations and RSP sea ice (ERA-Interim reanalysis treats the icebergs as sea ice) during the three iceberg calving events: B-9 (October 1987), B-15 (A,B,D) and B-17 (April 2000), and C-18 and C-19 (May 2002) (Martin et al., 2007). None of the calving events occur during DJF, thus iceberg impacts on DJF polynya area are expected in the following year's DJF sea ice, because of iceberg blocking effects on sea ice observed in the post years. Ba concentration linked with RSP sea ice fluctuations during these post years affect the biogenic bloom via sunlight blocking. Vertical lines in figure 3 denote calving events and arrows denote the affected year's response in SIC and subsequent RICE Ba concentrations. Detailed description of the iceberg calving RSP associations follows.

Iceberg B-9 (150 x 35 km) calved in October of 1987 (Keys et al., 1990) and there were no other major calving events until 1997 (Keys et al., 1998). This calving event is marked by a large peak in RICE Ba and DJF sea ice during 1988/1989 DJF (relatively small polynya), and the hiatus until 1997 is recorded as some of the lowest Ba and SIC over the satellite era (largest polynya area in the instrumental period). Icebergs B-15 (A, B, D) (300 x 40 km) and B-17 (100 x 20 km) calved north of Roosevelt Island in April of 2000, and slowly drifted through the RSP (Arrigo, 2003a), restricting summer sea ice breakout during 2000/2001 DJF. The highest levels of Ba and SIC since 1990 are synchronous with this period, yielding the smallest polynya since iceberg B-9 calved. During the 2001/2002 DJF season, B-15 was grounded off Ross Island, parallel to sea ice drift, west of the main RSP area with little effect on RSP productivity (Rhodes et al., 2009). RICE Ba levels synchronously return to average concentrations observed before B-15 and B-17 calving events during this period.

Icebergs C-18 (70 x 6 km) and C-19 (200 x 30 km) calved east of Ross Island in May of 2002 (Martin et al., 2007) and drifted north through the RSP, again restricting sea ice breakout (Arrigo, 2003a). Iceberg C-19 was positioned perpendicular to sea ice drift during 2002/2003 DJF, limiting RSP sea ice export greatly (Arrigo, 2003a) and the largest peak in the entire 1150 year record of Ba and SIC (smallest polynya) is recorded during 2002/2003 DJF. The lowest DJF levels of methylsulphonate (produced by biogenic activity in the RSP) in a study covering the period 1998-2004 (Rhodes et al., 2009) support our findings of decreased biogenic activity during this RSP sea ice year.

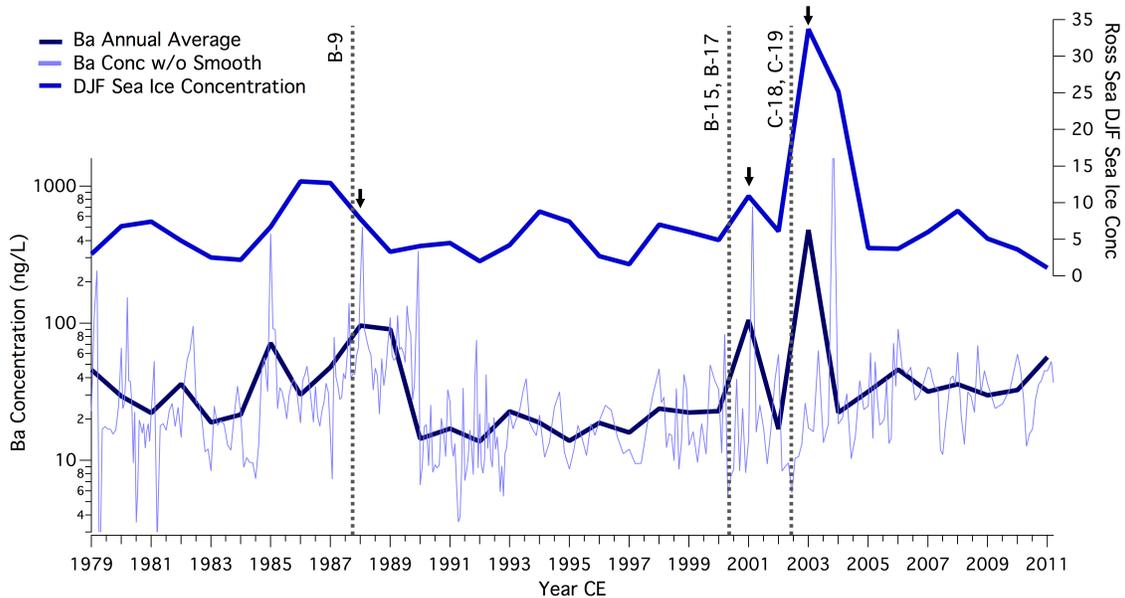

Figure 3: Annual average Ba concentration (with underlying raw data) plotted against DJF Ross Sea sea ice concentration using ERA Interim climate reanalysis data. Notice the sharp increases in sea ice and Ba concentrations during three iceberg calving events: B-9 (October 1987), B-15 (A,B,D) (April 2000), and C-18 and C19 (May 2002). None of the calving events occur during DJF, thus icebergs impact the following DJF sea ice and Ba concentrations that are associated with the austral summer biogenic bloom. The correlation is negative with biogenic blooms, and positively correlated with RSP sea ice via sunlight blocking (see text above for further explanation).

The 1150 year long RICE Ba record (figure 4) contains trends and large steps in concentration. The highest levels of Ba are recorded during the Modern Era (~1850 CE to present) climbing steadily out of the Little Ice Age (LIA). Ba levels during the LIA (assumed here to be 1400-1850 CE) are the second highest in the record. We use Siple Dome sea salt (ss) Na as a proxy for the Amundsen Sea low (ASL) (Kreutz et al., 2000) and Siple Dome non-sea-salt (nss) Ca as a proxy for South Hemisphere westerly (SHW) wind strength (Dixon et al., 2011; Yan et al., 2005) to provide context for other climatic variables over the record. The Medieval Warm Period (MWP) (assumed here to be 850 - 1400 CE) is marked by the lowest variability in Ba (RSP area and productivity) and in ASL depth and SHW wind strength over the last ~1150 years suggesting, based on our RSP study, that the MWP was the most climatically stable period in our record.

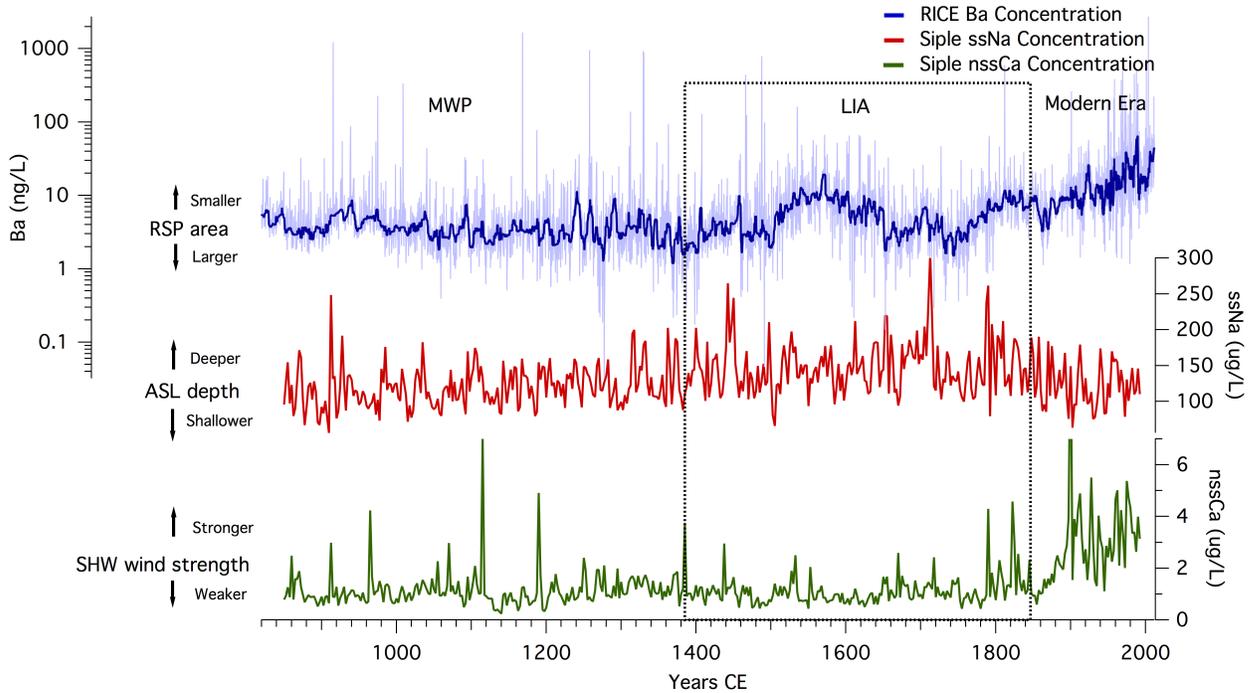

Figure 4: RICE Ba (proxy for RSP) and Siple Dome ssNa and nssCa (established proxies) over the last ~1150 years BP including the Medieval Warm Period, the Little Ice Age and the Modern Era. Siple Dome ssNa is a proxy for the ASL (Kreutz et al., 2000) and Siple Dome nssCa is a proxy for SHW wind strength (Dixon et al., 2011; Yan et al., 2005).

The LIA (1400 – 1850 CE) has a highly variable RSP signal characterized by stepwise fluctuations not observed during stable MWP conditions. RSP area undergoes a decline ~1350-1450 CE, then increases markedly from ~1450-1500 CE. ~1500 to 1625 CE is the most rapid prolonged decline in RSP area since the MWP, followed by RSP area returning to pre LIA levels from 1625-1750 CE, as supported by diatom assemblages, suggesting a more prevalent polynya (Leventer and Dunbar, 1988) with greater sea ice extent at this time. Persistent katabatic winds and cooler sea surface temperatures characterize the Ross Sea over the same period (Leventer and Dunbar, 1988), along with a deeper ASL(Kreutz et al., 2000) and increased ABW formation (Broecker et al., 1999), thus increasing Ross Sea $CO_2$ sequestration (Broecker et al., 1999; Etheridge et al., 1996; Indermühle et al., 1999). Based on our study the RSP expanded synchronously with these variables.

The stepwise fluctuations recorded in RICE Ba concentrations during the LIA may be an indication of the RSP's response to changes in Ross Sea storm frequency as noted in the Siple Dome ice core record (Kreutz et al., 1997). The Siple Dome ssNa record demonstrates high variability in ASL strength, but an overall deepening trend to present (Kreutz et al., 1997) likely associated with an increase in Ross Sea cyclogensis. Of specific note in figure 4 is the deepest relative ASL in the entire record (during the LIA at ~1450 and ~1700 CE) followed by brief periods of increased RSP area, implying increased katabatic flow and cyclone activity in the Ross Sea. The Law Dome $CO_2$ record shows a 6 ppm decline (Etheridge et al., 1996) from 1500-1800 CE during a period of increased polynya area inferred from our study. We speculate that $CO_2$ decline at this time is partially (albeit minimally) due to polynyas enabling increased deep-water ventilation.

Frequency of calving events during the LIA is unknown, but the effect is demonstrated during the instrumental period. Icebergs hinder sea ice export and block sunlight, both of which decrease biogenic activity during austral summers. Increased (decreased) calving events could be a partial explanation for smaller (larger) RSP area during the LIA. Potential increased calving during the LIA could be a result of colder conditions (Bertler et al., 2011), and a deeper ASL (Kreutz et al., 2000), causing an increase in Ross Ice Shelf mass balance.

RSP relative area decreases from 1750 to present, substantiated by the trend recorded in the instrumental period (supplementary figure 7), and now extended back in time by our proxy. Contraction of the polar vortex and associated contraction of the SHW (Bertler et al., 2011) over this period would result in pinning the ASL closer to coastal West Antarctica, and katabatic wind strength relative to LIA conditions remain constant (Bertler et al., 2011). SHW wind strength (plotted in figure 4) (Dixon et al., 2011; Yan et al., 2005) shows a marked increasing trend from 1750 CE to present and we suggest the decrease in RSP area over the Modern Era is likely a reflection of SHW strengthening and polar cell contraction overpowering katabatic flow in the RSP region, creating less offshore transport of sea ice. We observe the smallest RSP area of the entire 1150-year record in the Modern Era. The Law Dome $CO_2$ record also reflects this trend with a sharp increase starting at 1750 CE simultaneous with and partially (albeit minimally) explained in this study by decreased RSP area, followed by continued $CO_2$ concentration increases from 1850 CE to present, attributable to onset of the industrial period (Etheridge et al., 1996). Although RSP area decline is not the controlling factor in global atmospheric $CO_2$, our work suggests that Ross Sea $CO_2$ sequestration has decreased over the Modern Era, which could be driven, ironically by increasing atmospheric $CO_2$ through SHW contraction (Shulmeister et al., 2004) overpowering katabatic winds that are essential in forcing RSP sea ice offshore and therefore sustaining the polynya.

With the prediction of further contraction of the SHW due to increasing greenhouse warming (Shulmeister et al., 2004), our study infers that the RSP will continue its trend of decreasing area. As part of this loss of RSP area Ross Sea $CO_2$ sequestration would decrease because of declining ABW formation (less sea ice solidifying, less brine expelled into the ocean causing a density change) and declining ventilation (the smaller the polynya area exposed, the less sunlight to fuel a biogenic bloom, resulting in less carbon fixation in the Ross Sea). In the coming decades the RSP could disappear completely as the Weddell Sea Polynya has, as a result of anthropogenic forcing (Arrigo, 2003b), further reinforcing the effects of climate change in the Antarctic.

**Methods:**

We annual layer dated the RICE record using seasonal timing of sulfur (S), sodium (Na), iodine (I) and δ$^{18}$O. Prominent sulfate peaks coincide with the annual opening of the Ross Sea and biogenic activity during the summer season (Arrigo, 2003b; 2003a; Rhodes et al., 2009). The accumulation rate (~21 cm ice/yr) at RICE allowed a sampling resolution of ~11 inductively couple plasma mass spectrometer (ICP-MS) samples/yr allowing for ~4 samples/yr during the December-January-February (DJF) period over the ERA-Interim climate reanalysis period (1979-2012) used in calibrating our RSP record. ICP-MS samples were acidified to 1% nitric acid concentration with an internal standard, and analyzed 30 days after acidification; detection limits can be found in another study (Osterberg et al., 2006).

The upper 13 meters of the RICE record capture the instrumental era and annual layer counting is available to 1900 CE, after which methane dating correlated to the WAIS divide core and the WAIS divide 1251 CE volcanic tephra layer are used to extend dating to ~820 CE. Annual averages of ICP-MS generated concentrations (Sr, Cd, Cs, Ba, La, Ce, Pr, Pb, Bi, U, As, Li, I, Al, S, Ca, Ti, V, Cr, Mn, Fe, Co, Na, Mg, Cu, Zn, and K) were calculated, and empirical orthogonal function analysis (EOFs) preformed (Meeker et al., 1995). EOF1 is primarily sea salts (e.g., Na, Mg, Sr, V, Ca, K, explaining 27% of the total chemistry variance), EOF2 represents terrestrial dust components (e.g., Al, Ti, Pb, La, Ce, Li, Pr, 18% total chemistry variance), and EOF9 explains 32% of Ba variance and 3.5% of the total chemistry variance (see supplementary table 1). We then tested correlations of all element annual averages against ERA-Interim climate reanalysis variables (mean temperature 2m, sea surface temperature, U and V wind, wind speed, MSLP, precipitation, accumulated snow, sea ice concentration, total column water, total column ozone, total cloud cover, freezing degrees, melting degrees, snow mass balance and geo-potential height) and the strongest of the correlations were refined using sub-annual, max, min, and area under seasonal concentration peak calculations.

**Acknowledgements:**

This work is a contribution to the Roosevelt Island Climate Evolution (RICE) Program, funded by national contributions from New Zealand, Australia, Denmark, Germany, Italy, the People's Republic of China, Sweden, United Kingdom, and the United States of America. A special thanks to Ed Brooke, James Edward Lee, and TJ Fudge for their help in dating the RICE ice core. The main logistic support was provided by Antarctica New Zealand (K049) and the US Polar Program (I-209M) and this research was supported by NSF grants PLR-1042883, 1203640, and 11420007.

**Supplementary Figures:**

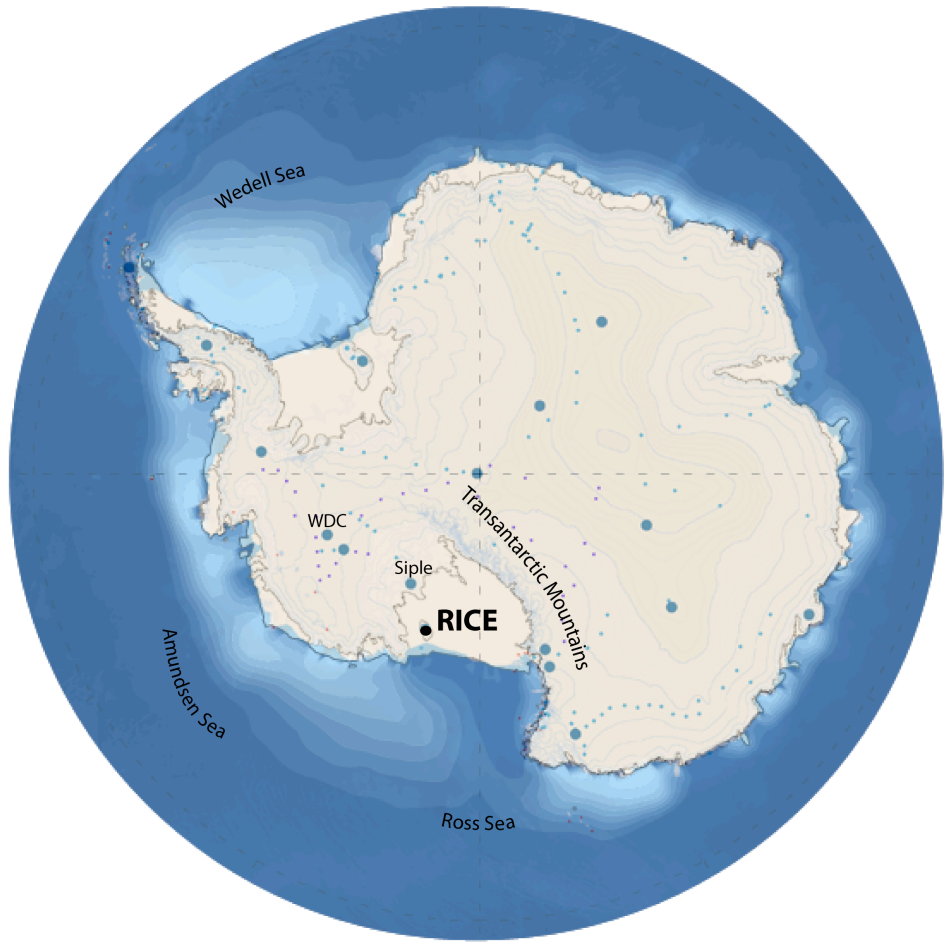

Supplementary figure 5: The RICE ice core was drilled to bedrock 120 km from the calving front of the Ross Ice Shelf, proximal to the Ross Sea Polynya and ~350 km coastward of the Siple Dome ice core and ~1000 km coastward of WAIS Divide ice cores. 1979-2011 DJF sea ice concentration is plotted to show the shape and location of the RSP. Note that the Transantarctic Mountains flank the Ross Sea Embayment offering sufficient topographic relief to produce the katabatic winds integral to polynya formation. Unlabeled dots are previous deep ice cores throughout Antarctica, note that RICE is one of the most coastal deep ice cores ever drilled, and that it is well positioned to capture RSP fluctuations.

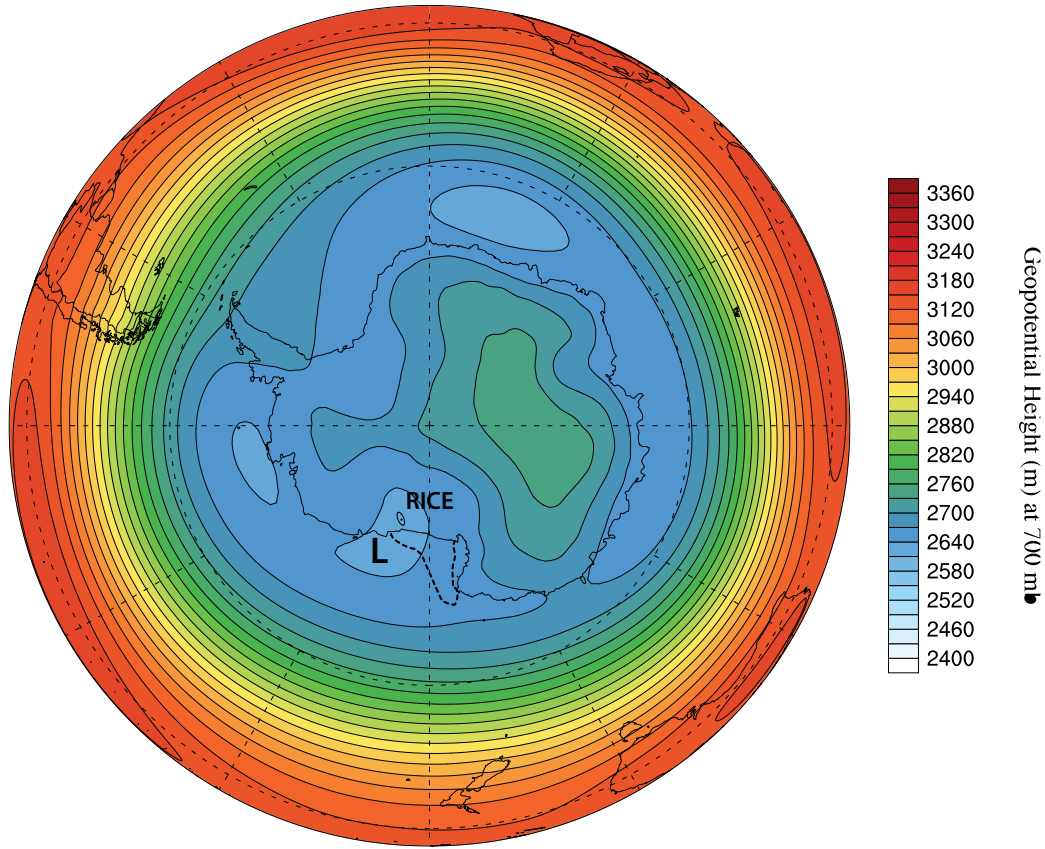

Supplementary figure 6: A persistent low-pressure system (denoted by "L") in the Ross Sea embayment east of the Ross Sea katabatics over DJF. This proximal low-pressure system has been shown by several authors (Bromwich et al., 1998; Drucker et al., 2011) to accelerate katabatic winds that are ultimately linked with polynya expansion during DJF. The low-pressure system acts simultaneously to entrain sea spray from the polynya (outlined with a dashed line) and deposit the sea spray at RICE. Image obtained using Climate Reanalyzer (http://cci-reanalyzer.org), Climate Change Institute, University of Maine, USA.

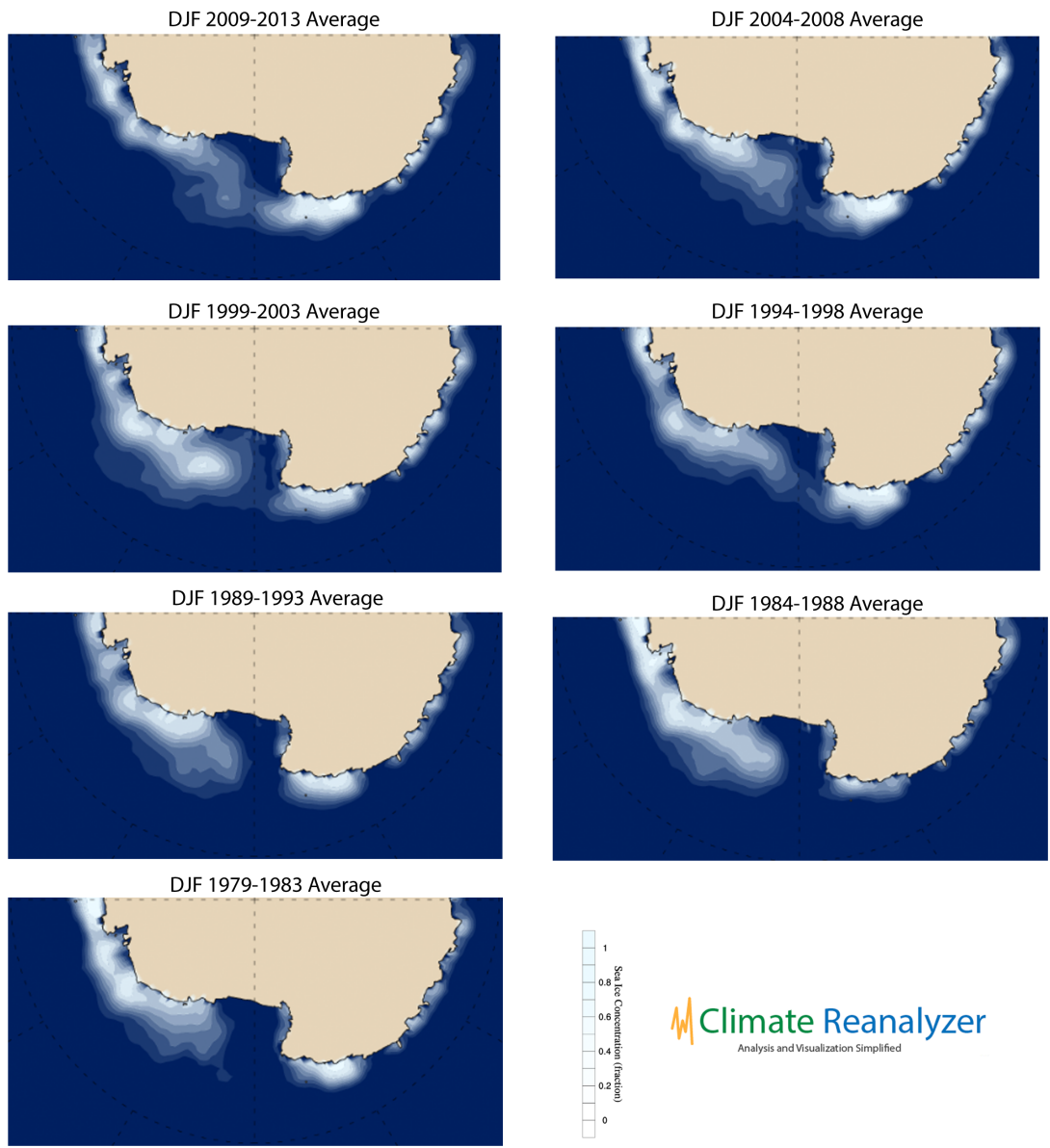

Supplementary figure 7: 5-year DJF sea ice averages from 1979-2013. Note the decreasing area of the RSP over the instrumental period as Ross Sea sea ice area increases. This matches the trend in Ba over this time period, even capturing the B-15 iceberg calving event over the 2001-2003. Image obtained using Climate Reanalyzer (http://cci-reanalyzer.org), Climate Change Institute, University of Maine, USA.

Supplementary table 1: RICE raw data empirical orthogonal function calculations after Meeker et al., 1995. Note that salts dominate EOF1, explaining 27% of the total variance of elements, EOF2 is dust components, explaining 18% of the total variance of elements, and EOF8 explains 32% of the variance of Ba. Total variance explained is denoted by T.V.E. in the table.

|        | EOF 1 | EOF 2 | EOF 3 | EOF 4 | EOF 5 | EOF 6 | EOF 7 | EOF 8 | EOF 9 |
|--------|-------|-------|-------|-------|-------|-------|-------|-------|-------|
| T.V.E. | 26.6  | 18.2  | 10.2  | 7.7   | 5.2   | 4.1   | 3.9   | 3.5   | 3.2   |
| δ18O   | -0.3  | 0.1   | -4.9  | -0.3  | -18.2 | 12.9  | 44.8  | -0.7  | -1.5  |
| Sr     | 53.8  | -6.4  | 30.7  | 3.8   | -0.7  | 0.0   | 1.2   | -0.1  | -0.4  |
| Cd     | 0.1   | 10.7  | -7.2  | 50.6  | -1.3  | -7.0  | -0.1  | 0.1   | -0.2  |
| Cs     | 49.6  | 7.8   | -19.1 | -0.1  | 0.3   | 0.1   | -1.7  | 0.8   | 0.0   |
| Ba     | 0.1   | 7.7   | -0.2  | 3.3   | -1.8  | -3.8  | 9.2   | 31.9  | 21.7  |
| La     | 1.4   | 85.5  | 8.2   | -2.1  | -1.1  | 0.0   | 0.0   | -0.1  | -0.1  |
| Ce     | 1.0   | 84.2  | 9.1   | -2.6  | -0.4  | 0.1   | 0.0   | -0.2  | -0.2  |
| Pr     | 1.5   | 83.9  | 8.7   | -2.5  | -1.1  | 0.1   | -0.1  | -0.2  | -0.1  |
| Pb     | 0.9   | 15.5  | -2.0  | 28.1  | 0.1   | 10.4  | -6.7  | 2.0   | -0.1  |
| Bi     | 0.6   | 1.0   | -0.2  | 4.1   | 1.3   | 56.0  | -4.9  | 7.1   | -2.5  |
| U      | 74.3  | 0.3   | -11.7 | -4.1  | 0.0   | 0.3   | -0.2  | 0.0   | 0.0   |
| As     | 13.0  | 0.0   | 0.5   | -0.4  | 8.1   | -0.3  | 6.2   | 31.2  | -2.5  |
| Li     | 41.1  | -6.3  | 36.6  | 9.9   | -0.4  | 0.1   | 0.6   | -0.1  | -0.2  |
| I      | 5.2   | -4.5  | 30.7  | 0.0   | 0.0   | -0.6  | -16.6 | 1.3   | 9.0   |
| Al     | 0.4   | 55.2  | 1.1   | 0.9   | -1.5  | -0.3  | 0.3   | 2.2   | 0.1   |
| S      | 75.8  | -0.4  | -14.5 | -3.5  | -0.4  | 0.0   | 0.2   | 0.0   | -0.1  |
| Ca     | 76.6  | -0.2  | -12.3 | -3.5  | -0.1  | -0.4  | -0.4  | -0.2  | 0.2   |
| Ti     | 0.4   | 72.5  | 10.0  | -6.6  | -2.3  | -0.2  | 0.0   | -1.3  | 0.0   |
| V      | 66.7  | 1.8   | -16.0 | -9.0  | -0.1  | -0.3  | -0.5  | -0.1  | 0.2   |
| Cr     | 2.6   | 2.0   | -0.8  | 6.4   | 1.5   | 14.3  | 2.1   | -11.0 | 41.1  |
| Mn     | 9.4   | 25.5  | -0.1  | 0.0   | 35.6  | -0.1  | 1.1   | -1.6  | -0.2  |
| Fe     | 0.3   | 4.2   | 0.6   | 0.5   | 66.9  | -0.2  | 7.1   | -0.7  | -0.3  |
| Co     | 1.2   | 5.3   | -3.3  | 28.9  | 0.0   | -3.9  | 0.7   | -4.6  | 0.4   |
| Na     | 79.9  | -7.6  | 8.6   | 0.5   | -0.9  | 0.0   | 0.2   | -0.1  | -0.1  |
| Mg     | 64.3  | -7.5  | 19.8  | 2.0   | 0.0   | 0.0   | 1.5   | -0.4  | -0.3  |
| Cu     | 3.5   | 6.6   | -10.9 | 39.4  | -1.7  | -2.9  | -0.3  | -0.9  | -7.3  |
| Zn     | 39.1  | 0.9   | -13.1 | -1.9  | -0.1  | -0.2  | -1.3  | 0.1   | 1.2   |
| K      | 82.9  | -5.1  | 5.7   | 1.3   | -0.2  | 0.1   | 0.0   | -0.2  | 0.0   |